\documentclass[12pt,pdftex]{iopart}

\usepackage{lineno}
\usepackage{iopams}             
\usepackage{acronym}
\usepackage{graphicx}
\usepackage{caption}
\usepackage{subcaption}
\usepackage[colorlinks,pagebackref,breaklinks]{hyperref} 
\hypersetup{hypertexnames=true,
  linkcolor=blue,anchorcolor=black,citecolor=blue,urlcolor=blue
}
\DeclareGraphicsExtensions{.jpg,.pdf,.mps,.png}
\graphicspath{{imglocal/}{img/}{./}} 

\newcommand{\revised}[2]{#2} 

\acrodef{SW}[SW]{Software}
\acrodef{HW}[HW]{Hardware} 
\acrodef{IO}[I/O]{Input/Output}
\acrodef{AFM}[AFM]{Atomic Force Microscope} 
\acrodef{PCB}[PCB]{Printed Circuit Board} 
\acrodef{ICT}[ICT]{Information and Communication Technology} 
\acrodef{CMOS}[CMOS]{Complementary Metal Oxide Semiconductor} 
\acrodef{VLSI}[VLSI]{Very Large Scale Integration} 
\acrodef{SOI}[SOI]{Silicon on Insulator} 
\acrodef{MOSFET}[MOSFET]{Meal Oxide Semiconductor Field Effect Transistor} 
\acrodef{FET}[FET]{Field Effect Transistor} 
\acrodef{SRAM}[SRAM]{Static Random Access Memory}
\acrodef{DRAM}[DRAM]{Dynamic Random Access Memory}
\acrodef{RRAM}[ReRAM]{Resistive Random Access Memory}
\acrodef{CPU}[CPU]{Central Processing Unit}
\acrodef{FSM}[FSM]{Finite State Machine}
\acrodef{AER}[AER]{Address Event Representation}
\acrodef{WTA}[WTA]{Winner-Take-All}
\acrodef{IF}[I\&F]{Integrate-and-Fire}
\acrodef{NMDA}[NMDA]{N-Methyl-D-Aspartate}
\acrodef{LTP}[LTP]{Long Term Potentiation}
\acrodef{LTD}[LTD]{Long Term Depression}
\acrodef{DAC}[DAC]{Digital to Analog Converter}
\acrodef{ADC}[ADC]{Analog to Digital Converter}
\acrodef{FPGA}[FPGA]{Field Programmable Gate Array}
\acrodef{SRAM}[SRAM]{Static Random Access Memory}
\acrodef{DPI}[DPI]{Differential Pair Integrator}
\acrodef{STDP}[STDP]{Spike-Timing Dependent Plasticity}
\acrodef{PSTH}[PSTH]{Peri-Stimulus Time Histogram}
\acrodef{EPSP}[EPSP]{Excitatory Post Synaptic Potential}
\acrodef{EPSC}[EPSC]{Excitatory Post Synaptic Current}
\acrodef{IPSC}[IPSC]{Inhibitory Post Synaptic Current}
\acrodef{ISI}[ISI]{Inter-Spike Interval}
\acrodef{STD}[STD]{Short-Term Depression}
\acrodef{HH}[H\&H]{Hodgkin-Huxley}
\acrodef{RNN}[RNN]{Recurrent Neural Network}
\acrodef{ANN}[ANN]{Attractor Neural Network}
\acrodef{CMOL}[CMOL]{``Hybrid CMOS nanoelectronic circuits''}
\acrodef{MIM}[MIM]{Metal Insulator Metal}
\acrodef{EBL}[EBL]{Electron Beam Lithography}
\acrodef{NIL}[NIL]{Nano-Imprint Lithography}
\acrodef{MCMC}[MCMC]{Markov-Chain Monte Carlo}
\acrodef{HRS}[HRS]{High-Resistive State}
\acrodef{LRS}[LRS]{Low-Resistive State}

\begin{document}

\title[Neuromorphic nanoscale memristor synapses]{Integration of
  nanoscale memristor synapses in neuromorphic computing
  architectures}

\author{Giacomo Indiveri$^{a,1}$, Bernabe Linares-Barranco$^b$, Robert Legenstein$^c$,
  George Deligeorgis$^d$, and Themistoklis Prodromakis$^e$}

\address{$^a$ Institute of Neuroinformatics, University of Zurich and ETH Zurich, Zurich, Switzerland\\
$^b$Instituto Microelectronica Sevilla  (IMSE-CNM-CSIC), Sevilla, Spain\\
$^c$Institute for Theoretical Computer Science, Graz University of Technology, Graz, Austria\\
$^d$CNRS-LAAS and Univ de Toulouse, 7 avenue du colonel Roche, F-31400 Toulouse, France\\
$^e$Center for Bio-Inspired Technology, Department of Electrical and Electronic Engineering, Imperial College London, United Kingdom
}
\ead{$^1$giacomo@ethz.ch}
\begin{abstract}
  Conventional neuro-computing architectures and artificial neural
  networks have often been developed with no or loose connections to
  neuroscience. As a consequence, they have largely ignored key
  features of biological neural processing systems, such as
  \revised{low-power}{their extremely low-power consumption features
    or their} ability to carry out robust and efficient computation
  using \revised{parallel}{massively parallel arrays of} limited
  precision, highly variable, and unreliable components.  Recent
  developments in nano-technologies are making available extremely
  compact and low-power, but also variable and unreliable solid-state
  devices that can potentially extend the offerings of availing \acs{CMOS}
  technologies.  In particular, memristors are regarded as a promising
  solution for modeling key features of biological synapses due to
  their nanoscale dimensions, their capacity to store multiple bits of
  information per element and the low energy required to write
  distinct states.  In this paper, \revised{review}{we first review
    the neuro- and neuromorphic-computing approaches that can best
    exploit the properties of memristor and-scale devices, and
    then} propose a novel hybrid memristor-\acs{CMOS} neuromorphic circuit
  \revised{radical}{which represents a radical departure from
    conventional neuro-computing approaches, as it uses memristors to
    directly emulate the biophysics and temporal dynamics of real
    synapses. We point out the differences between the use of
    memristors in conventional neuro-computing architectures and the
    hybrid memristor-\acs{CMOS} circuit proposed, and argue how this circuit
    represents an ideal building block} for implementing
  brain-inspired probabilistic computing paradigms that are robust to
  variability and fault-tolerant by design.
\end{abstract}


\maketitle

\section{Introduction}

The idea of linking the type of information processing that takes
place in the brain with theories of computation and computer science
(something commonly referred to as \emph{neuro-computing}) dates back
to the origins of computer science itself~\cite{McCulloch_Pitts43,
  Neumann58}. Neuro-computing has been very popular in the
past~\cite{Rosenblatt58,Minsky67}, eventually leading to the
development of abstract artificial neural networks implemented on
digital computers, useful for solving a wide variety of practical
problems~\cite{Hopfield82, Rumelhart_McClelland86, Kohonen88,
  Hertz_etal91, Bishop06}.  However, the field of \emph{neuromorphic
  engineering} is a much younger one~\cite{Indiveri_Horiuchi11}. This
field has been mainly concerned with hardware implementations of
neural processing and sensory-motor systems built using \ac{VLSI}
electronic circuits that exploit the physics of silicon to reproduce
directly the biophysical processes that underlie neural computation in
real neural systems.  Originally, the term ``neuromorphic'' (coined by
Carver Mead in 1990~\cite{Mead90}) was used to describe systems
comprising analog integrated circuits, fabricated using standard
\ac{CMOS} processes. In recent times, however, the use of this term
has been extended to refer to hybrid analog/digital electronic
systems, built using different types of technologies.

Indeed, both artificial neural networks and neuromorphic computing
architectures are now receiving renewed attention thanks to progress
in \acp{ICT} and to the advent of new promising nanotechnologies. Some
of present day neuro-computing approaches attempt to model the fine
details of neural computation using standard technologies. For
example, the \emph{Blue Brain} project, launched in 2005, makes use of
a 126kW Blue Gene/P IBM supercomputer to run software that simulates
with great biological accuracy the operations of neurons and synapses
of a rat neocortical column~\cite{Markram06}.  Similarly, the
\emph{BrainScaleS} EU-FET FP7 project aims to develop a custom neural
supercomputer by integrating standard \ac{CMOS} analog and digital
\ac{VLSI} circuits on full silicon wafers to implement about 262
thousand \ac{IF} neurons and 67 million
synapses~\cite{Schemmel_etal08}. Although configurable, the neuron and
synapse models are hardwired in the silicon wafers, and the hardware
operates about 10000 times faster than real biology, with each wafer
consuming about 1kW power, excluding all external components.  Another
large-scale neuro-computing project based on conventional technology
is the \emph{SpiNNaker} project~\cite{Jin_etal10}. The SpiNNaker is a
distributed computer, which interconnects conventional multiple
integer precision multi ARM core chips via a custom communication
framework. Each SpiNNaker package contains a chip with 18 ARM9
\acp{CPU} on it, and a memory chip of 128\,Mbyte Synchronous
\ac{DRAM}. Each \ac{CPU} can simulate different neuron and synapse
models. If endowed with simple synapse models, a single SpiNNaker
device \revised{ARM}{ARM core} can simulate the activity of about 1000
neuron in real time. More complex synapse models (e.g.\ with learning
mechanisms) would use up more resources and decrease the number of
neurons that could be simulated in real-time. The latest SpiNNaker
board contains 47 of these packages, and the aim is to assemble 1200
of these boards. A full SpiNNaker system of this size would consume
about 90\,kW.  The implementation of custom large-scale spiking neural
network hardware simulation engines is being investigated also by
industrial research groups. For example, the IBM group led by
D.S. Modha recently proposed a digital ``neurosynaptic core'' chip
integrated using a standard 45\,nm \ac{SOI}
process~\cite{Arthur_etal12}.  The chip comprises 256 digital \ac{IF}
neurons, with $1024\times 256$ binary valued synapses, configured via
a \ac{SRAM} cross-bar array, and uses an asynchronous event-driven
design to route spikes from neurons to synapses. The goal is to
eventually integrate many of these cores onto a single chip, and to
assemble many multi-core chips together, to simulate networks of
simplified spiking neurons with human-brain dimensions (i.e.\
approximately $10^{10}$ neurons and $10^{14}$ synapses) in
real-time. In the mean time, IBM simulated 2.084 billion neurosynaptic
cores containing $53\times 10^{10}$ neurons and $1.37\times 10^{14}$
synapses in software on the Lawrence Livermore National Lab Sequoia
supercomputer (96 Blue Gene/Q racks), running 1542$\times$ slower than
real time~\cite{Wong_etal12}, and dissipating 7.9\,MW\@. A
diametrically opposite approach is represented by the \emph{Neurogrid}
system~\cite{Silver_etal07}. This system comprises an array of sixteen
$12\times 14$\,mm$^2$ chips, each integrating mixed analog
neuromorphic neuron and synapse circuits with digital asynchronous
event routing logic. The chips are assembled on a $16.5\times
19$\,cm$^2$ \ac{PCB}, and the whole system can model over one million
neurons connected by billions of synapses in real-time, and using only
about 3\,W of power~\cite{Choudhary_etal12}. As opposed to the
neuro-computing approaches that are mainly concerned with fast and
large simulations of spiking neural networks, the Neurogrid has been
designed following the original neuromorphic approach, exploiting the
characteristics of \ac{CMOS} \ac{VLSI} technology to directly emulate
the biophysics and the connectivity of cortical circuits. In
particular, the Neurogrid network topology is structured by the data
and results obtained from neuro-anatomical studies of the mammalian
cortex. While offering less flexibility in terms of connectivity
patterns and types of synapse/neuron models that can be implemented,
the Neurogrid is much more compact and dissipates orders of magnitude
less power than the other neuro-computing approaches described above.
All these approaches have in common the goal of attempting to simulate
large numbers of neurons, or as in the case of Neurogrid, to physically
emulate them with fine detail.  

Irrespective of the approach followed, nanoscale synapse technologies
and devices have the potential to greatly improve circuit integration
densities and to substantially reduce power-dissipation in these
systems.  Indeed, recent trends in nanoelectronics have been
investigating emerging low-power nanoscale devices for extending
standard \ac{CMOS} technologies beyond the current
state-of-art~\cite{di2011memory}. In particular, \ac{RRAM} is regarded
as a promising technology for establishing next-generation
non-volatile memory cells~\cite{Rozenberg:2004ck}, due to their
infinitesimal dimensions, their capacity to store multiple bits of
information per element and the minuscule energy required to write
distinct states.  The factors driving this growth are attributed to
the devices' simple (two terminals) and infinitesimal structure
(state-of-art is down to 10$\times 10\,nm^2$~\cite{Govoreanu:2011ij})
and ultra-low power consumption ($<50$\,fJ/bit) that so far are
unmatched by conventional VLSI circuits.

Various proposals have already been made for leveraging basic
nanoscale \ac{RRAM} attributes in reconfigurable
architectures~\cite{Snider:2007ba},
neuro-computing~\cite{Hasegawa:2010jq} and even artificial
synapses~\cite{Serrano-Gotarredona_etal13,zamarreno2011spike,Jo:2010hh,Choi:2009eu,Kuzum:2012ge}.
However the greatest potential of these nanoscale devices lies in the
wide range of interesting physical properties they possess.
Neuromorphic systems can harness the interesting physics being
discovered in these new nanodevices to emulate the biophysics of real
synapses and neurons and reproduce relevant computational primitives,
such as state-dependent conductance changes, multi-level stability and
stochastic state changes in large-scale artificial neural systems.

In this paper we \revised{novel}{first describe how nanoscale synaptic
  devices can be integrated into neuro-computing architectures to
  build large-scale neural networks, and then propose a new hybrid
  memristor-\ac{CMOS} neuromorphic circuit that emulates the behavior
  of real synapses, including their temporal dynamics aspects}, for
exploring and understanding the principles of neural computation and
eventually building brain-inspired computing systems.

\section{Solid-state memristors}
\label{sec:memristors}

\ac{RRAM} cells are nowadays classified as being
memory-resistors~\cite{Chua:2011el}, or memristors for short, that
have first been conceptually conceived in 1971 by Leon
Chua~\cite{Chua:ce}; with the first biomimetic applications presented
at the same time. The functional signature of memristors is a pinched
hysteresis loop in the current-voltage (i-v) domain when excited by a
bipolar periodic stimulus~\cite{di2009circuit}. Such hysteresis is
typically noticed for all kind of devices/materials in support of a
discharge phenomenon that possess certain inertia, causing the value
of a physical property to lag behind changes in the mechanism causing
it, and has been common both to large scale~\cite{Prodromakis:2012uf}
as well as nanoscale dissipative devices~\cite{Snider:2009p329}.

\begin{figure}
  \centering
  \begin{subfigure}[b]{0.4\textwidth}
    \includegraphics[width=\textwidth]{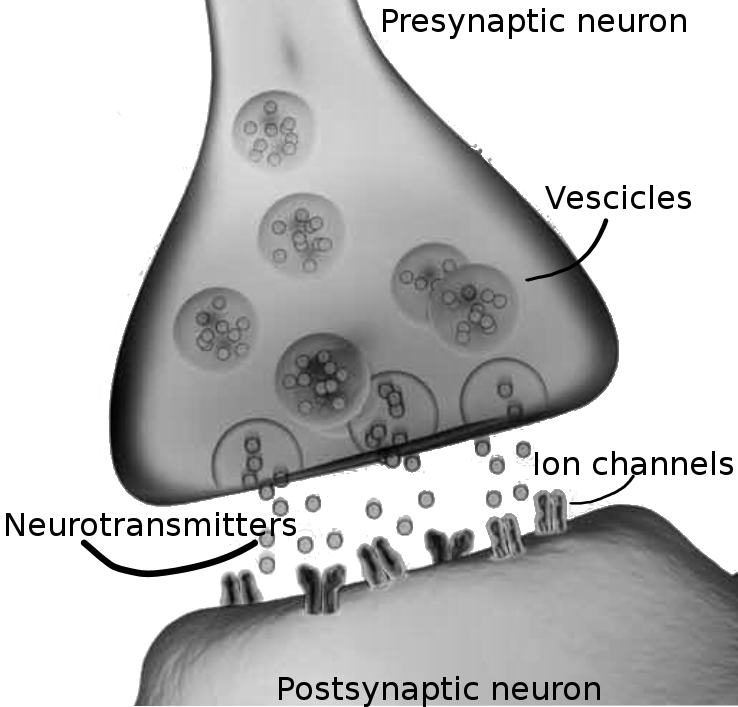}
    \caption{}
    \label{fig:synapse}
  \end{subfigure}
  \hfill
  \begin{subfigure}[b]{0.4\textwidth}
    \includegraphics[width=\textwidth]{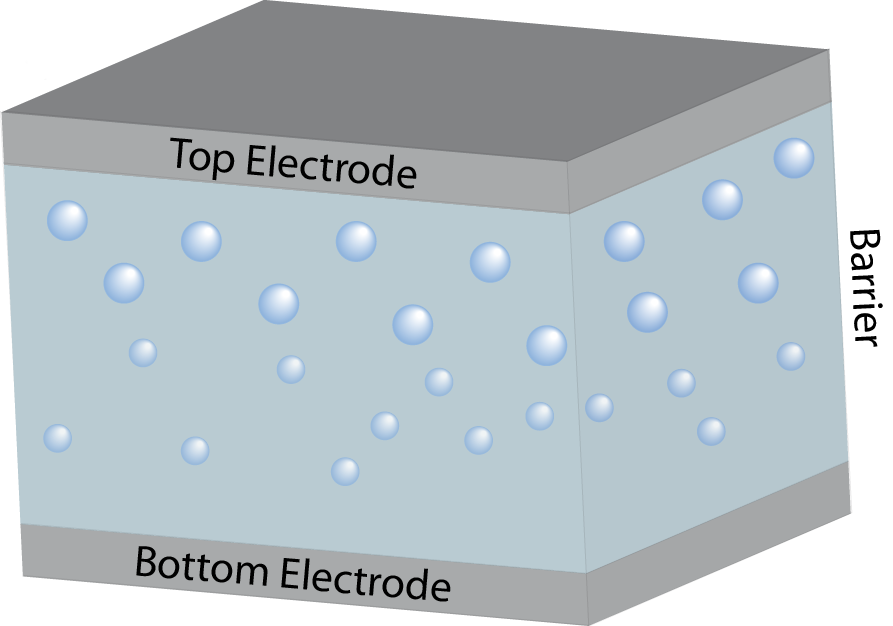}
    \caption{}
    \label{fig:memristor}
  \end{subfigure}
 \caption{(a) Cross-section of a chemical synapse, illustrating the
   discharge of neurotransmitters within a synaptic cleft originating
   from a pre-synaptic neuron. (b) Schematic representation of
   solid-state memristors where ionic species can be displaced within
   a device's insulating medium, transcribing distinct resistive
   states, by application of electrical stimuli on the top or bottom
   electrodes of the device.}
    \label{fig:mem-synapse}
\end{figure}

\subsection{Emerging nanodevices as synapse mimetics}
\label{sec:memr-devices}

The analogy of memristors and chemical synapses is made on the basis
that synaptic dynamics depend upon the discharge of neurotransmitters
within a synaptic cleft (see Fig.~\ref{fig:synapse}), in a similar
fashion that ``ionic species'' can be displaced within any inorganic
barrier (see Fig.~\ref{fig:memristor}).  $TiO_2$-based memristor
models~\cite{Snider:2009p329,Prodromakis:ge} hypothesized that
solid-state devices comprise a mixture of $TiO_2$ phases, a
stoichiometric and a reduced one ($TiO_2-x$), that can facilitate
distinct resistive states via controlling the displacement of oxygen
vacancies and thus the extent of the two phases. More recently however
it was demonstrated that substantial resistive switching is only
viable through the formation and annihilation of continuous conductive
percolation channels~\cite{Shihong:2012ul} that extend across the
whole active region of a device, shorting the top and bottom
electrodes; no matter what the underlying physical mechanism is.

\begin{figure}
  \centering
  \begin{subfigure}[b]{0.425\textwidth}
    \includegraphics[width=\textwidth]{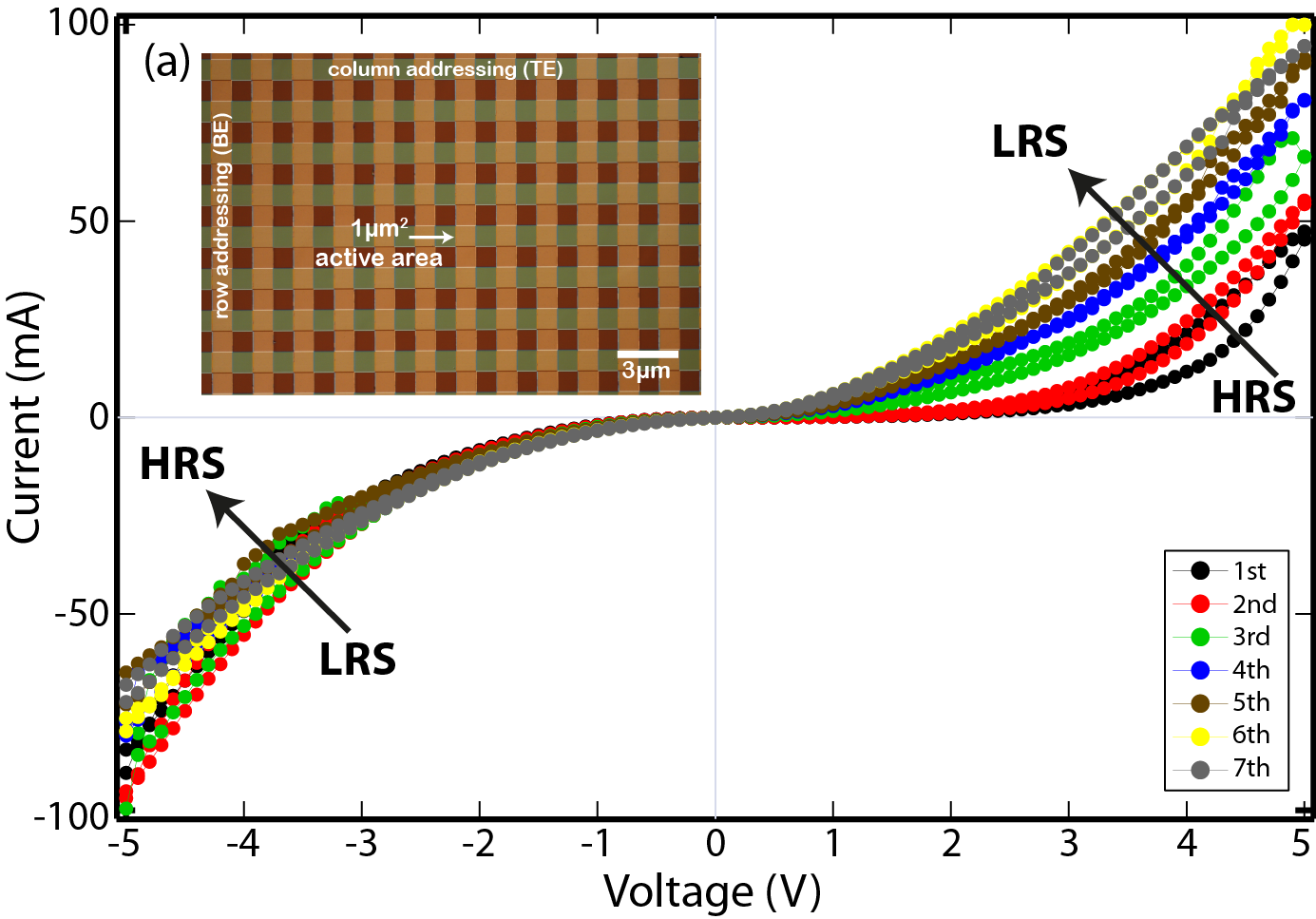}
    \caption{}
    \label{fig:memiv}
  \end{subfigure}\hfill
  \begin{subfigure}[b]{0.5\textwidth}
    \includegraphics[width=\textwidth]{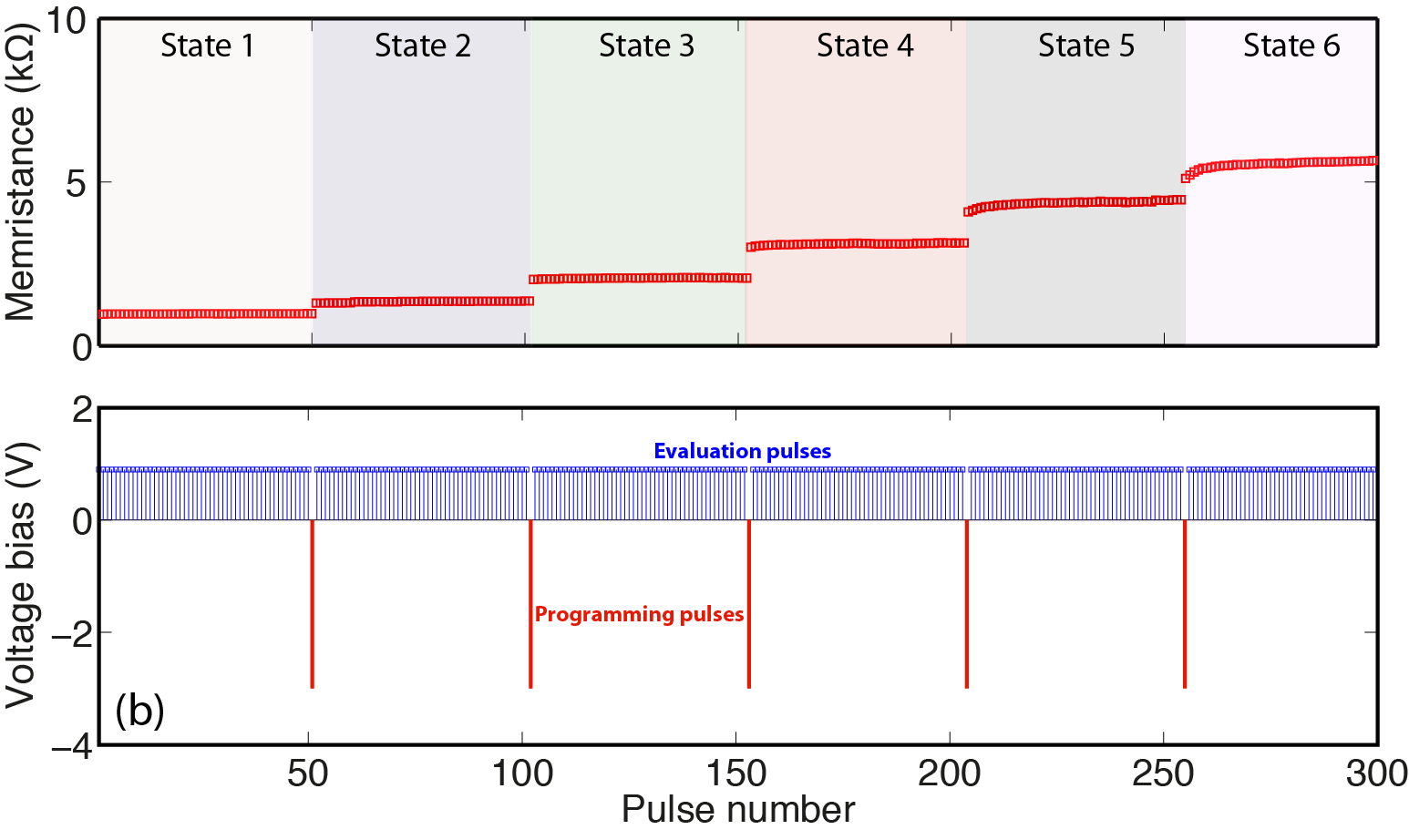}
    \caption{}
    \label{fig:mempulsing}
  \end{subfigure}
  \caption{Characterization of a $TiO_2$-based solid-state
   memristor. (a) I-V characteristics for consecutive voltage
   sweeping. Positive (negative) biasing renders an increase
   (decrease) in the device's conductance. Inset of (a) depicts a
   $25\times 25$ array crossbar type memristors comprising of $TiO_2$
   active areas of $1\times 1\mu m^2$. These cells can be programmed
   at distinct resistive states as shown in (b) by employing -3V and
   $1\mu sec$ wide pulses, while evaluation of the device's states is
   performed at 0.9V.}
    \label{fig:mem-data}
\end{figure}

An example of I-V characteristics of $TiO_2$-based memristors is
depicted in Fig.~\ref{fig:memiv}. In this example, consecutive
positive voltage sweeps cause any of the cross-bar type
devices~\cite{Kim_etal12} shown in the inset of Fig.~\ref{fig:memiv}
to switch from a \ac{HRS} to \acp{LRS}. When the polarity of the
voltage sweeps is however inverted, the opposite trend occurs, i.e.\
the device toggles from \ac{LRS} to \ac{HRS} consecutively (as
indicated by the corresponding arrows). These measured results
\revised{multi-level}{are consistent with analogous ones proposed by
  other research groups~\cite{Liu_etal09,Yoon_etal12,Yang_Chen12} and}
demonstrate the devices' capacity for storing a multitude of resistive
states per unit cell, with the programming depending on the biasing
history. This is further demonstrated in Fig.~\ref{fig:mempulsing}, by
applying individual pulses of -3\,V in amplitude and 1\,$\mu$sec long
for programming a single memristor at distinct non-volatile resistive
states. In this scenario, the solid-state memristor emulates the
behavior of a depressing
synapse~\cite{ODonovan_Rinzel97,Chance_etal98}; the inverse, i.e.\
short-term potentiation is also achievable by alternating the polarity
of the employed pulsing scheme.

The development of nanoscale dynamic computation elements may notably
benefit the establishment of neuromorphic architectures. This
technology adds substantially on computation functionality, due to the
rate-dependency of the underlying physical switching mechanisms. At
the same time it can facilitate unprecedented complexity due to the
capacity of storing and processing spiking events locally. Moreover,
exploiting the nanoscale dimensions and architecture simplicity of
solid-state memristor implementations could substantially augment the
number of cells per unit area, effectively enhancing the systems'
redundancy for tolerating issues that could stem from device mismatch
and low-yields~\cite{Gelencser:2012te}.

\subsection{Memristor scaling}
\label{sec:scaling}


Resistive memory scaling has been intensively investigated for
realization of nanosized \ac{RRAM}~\cite{yang_linear_2012}. In
principle memristors may be scaled aggressively well below
conventional RAM cells due to their simplicity: fabrication-wise
memristors typically rely on a \ac{MIM} structure. The memristor
action occurs in the insulating material. Scaling down the thickness
of such a material will reduce both the required set voltage as well
as the read voltage used during operation. In this context, thickness
figures of a few nano meters have been demonstrated and operating
voltages below 1\,V have been shown~\cite{deleruyelle_2013} with a
switching energy of a few fJ~\cite{chin_ultra-low_2012}.  Furthermore,
reducing the active device area by down-scaling the electrodes leads
to current scaling, as well as increased device density. Both of these
effects are favorable for high complexity circuits.

\begin{figure}
  \centering
  \includegraphics[width=0.65\textwidth]{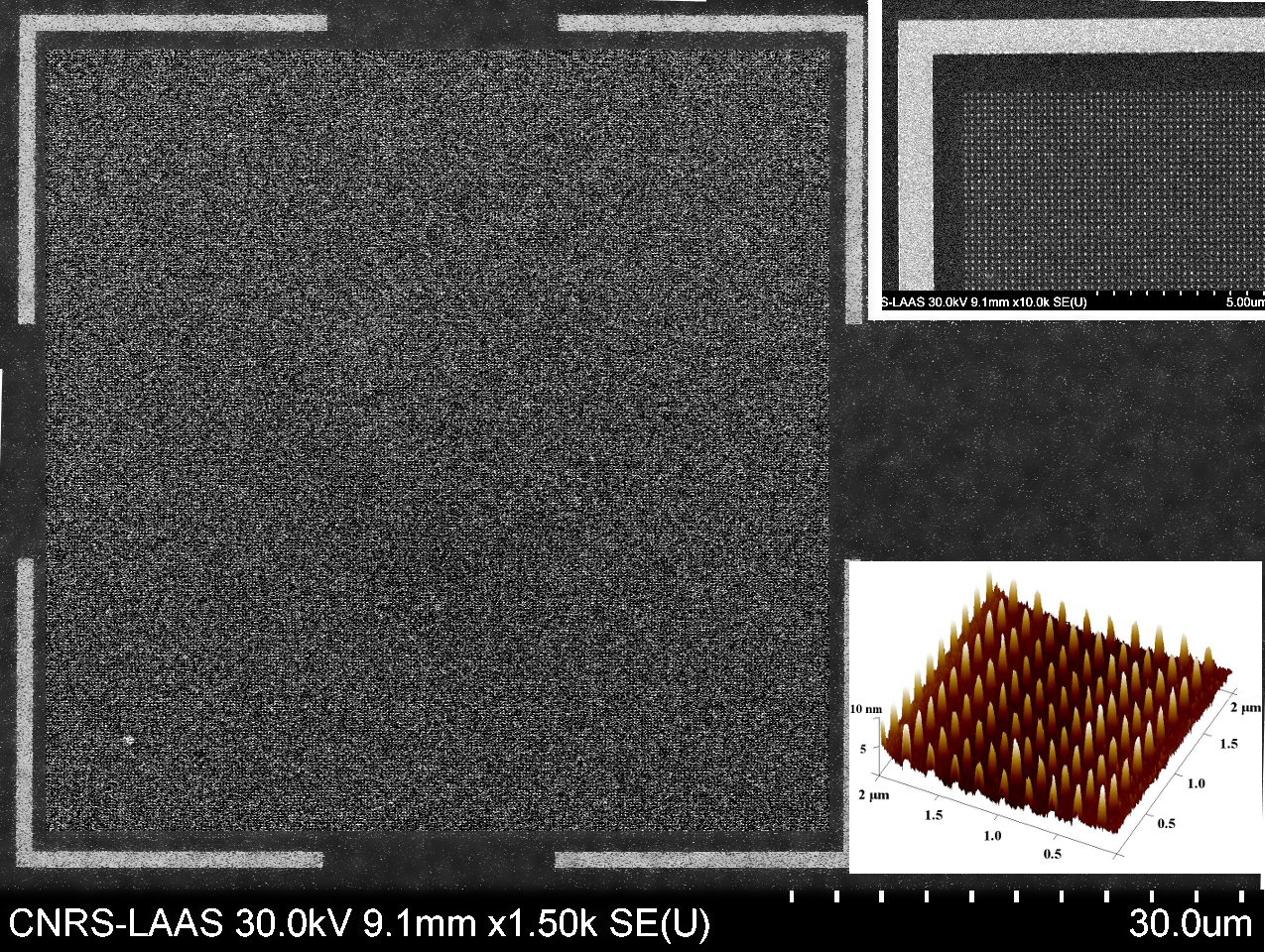}
  \caption{SEM micrograph of a large nanosized memristor array. Top
    inset shows a zoom-in of the top left corner where the individual
    devices are distinguished. Bottom left inset shows an \ac{AFM} image of
    a small part of the array. Individual devices are addressed by
    placing a conductive \ac{AFM} tip on the top electrode.}
  \label{fig:sem-micrograph}
\end{figure}

Currently even though single memristor devices as small as $10\times
10$\,nm have been demonstrated~\cite{Govoreanu:2011ij}, cross-bar
arrays are the most commonly used
architecture~\cite{lewis_architectural_2009,Kim_etal12} to organize large numbers
of individually addressable memristive synapses in a reduced space.
In Fig.~\ref{fig:sem-micrograph} we show a large array of nanoscale
memristors that we fabricated using electron beam lithography. This
array consists of a continuous Pt bottom electrode and an active layer
deposited by Sputtering. Subsequently, several arrays of
nano-memristors with a size ranging from 20 to 50\,nm were defined using
E-beam lithography on PMMA and lift-off of the top Platinum
electrode. The array shown here comprises $256 \times 256$ devices
with a periodicity of 200\,nm. To access each individual device a
conductive \ac{AFM} tip was used. Such a structure has been used to
study the variability of the fabricated devices. Using E-beam
lithography for both the top and bottom electrodes a fully
interconnected cross bar structure with similar size and pitch may be
fabricated.

\section{Memristor-based neuro-computing architectures}
\label{sec:cmos-nano}

Memristive devices have been proposed as analogs of biological
synapses. Indeed, memristors could implement very compact but abstract
models of synapses, for example representing a binary ``potentiated''
or ``depressed'' state, or storing an analog ``synaptic weight''
value~\cite{kudithipudi_reconfigurable_2012}. In this framework, they
could be integrated in large and dense cross-bar
arrays~\cite{likharev_crossnets:_2011} to connect large numbers of
silicon neurons~\cite{Indiveri_etal11}, and used in a way to implement
spike-based learning mechanisms that change their local conductance.

\begin{figure}
  \centering
  \begin{subfigure}[b]{0.55\textwidth}
    \includegraphics[width=\textwidth]{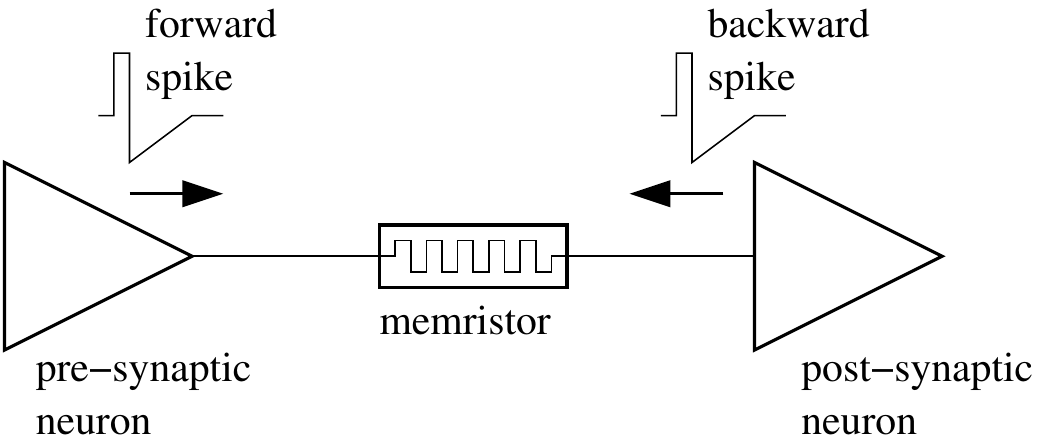}
    \caption{}
    \label{fig:onesynapse}
  \end{subfigure}
  \begin{subfigure}[b]{0.35\textwidth}
    \includegraphics[width=\textwidth]{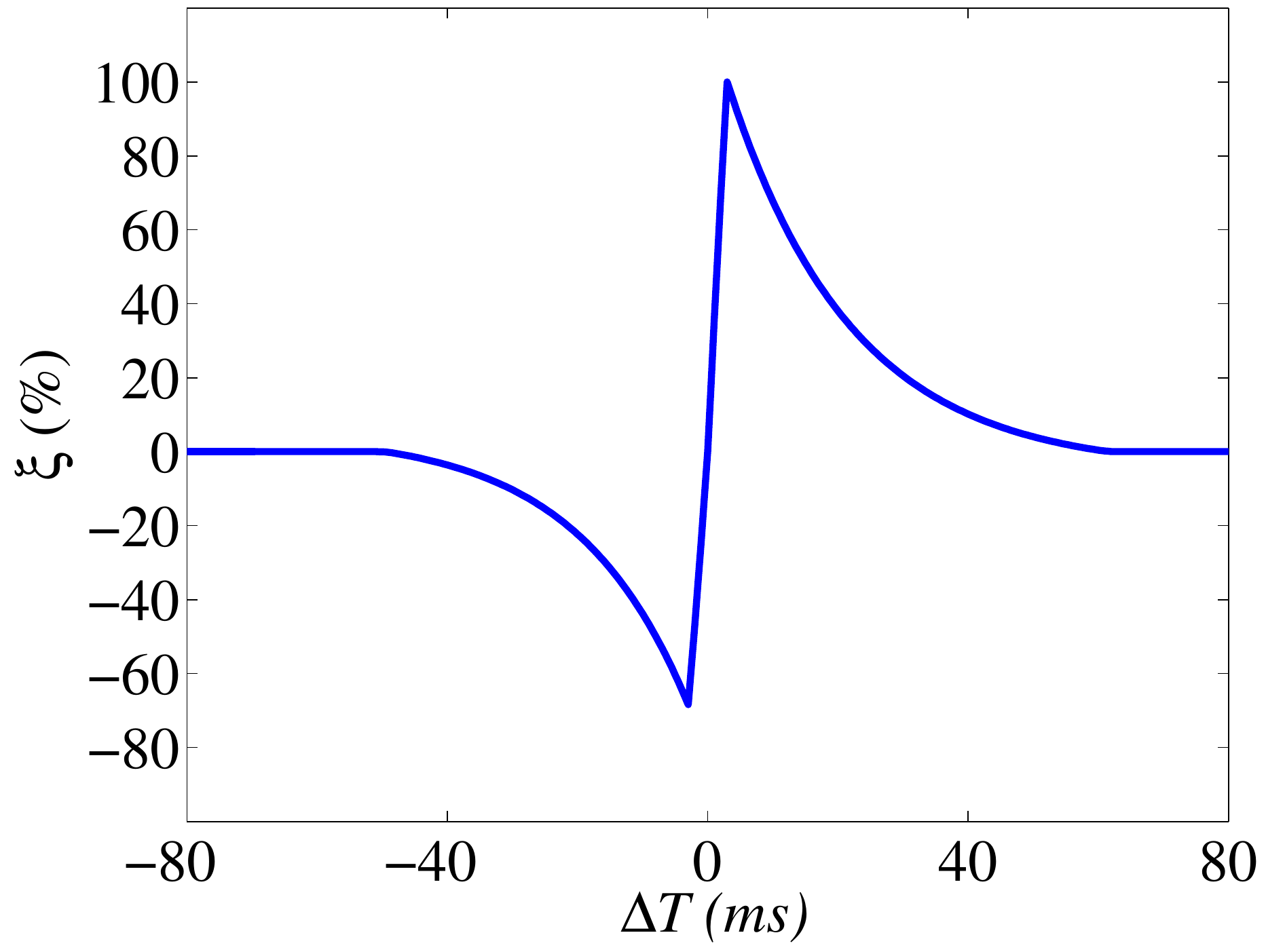}
    \caption{}
    \label{fig:stdpsim}
  \end{subfigure}\\
  \begin{subfigure}[b]{0.5\textwidth}
    \includegraphics[width=\textwidth]{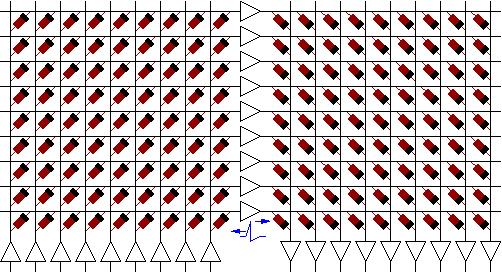}
    \caption{}
    \label{fig:xbar}
  \end{subfigure}
  \begin{subfigure}[b]{0.45\textwidth}
    \includegraphics[width=\textwidth]{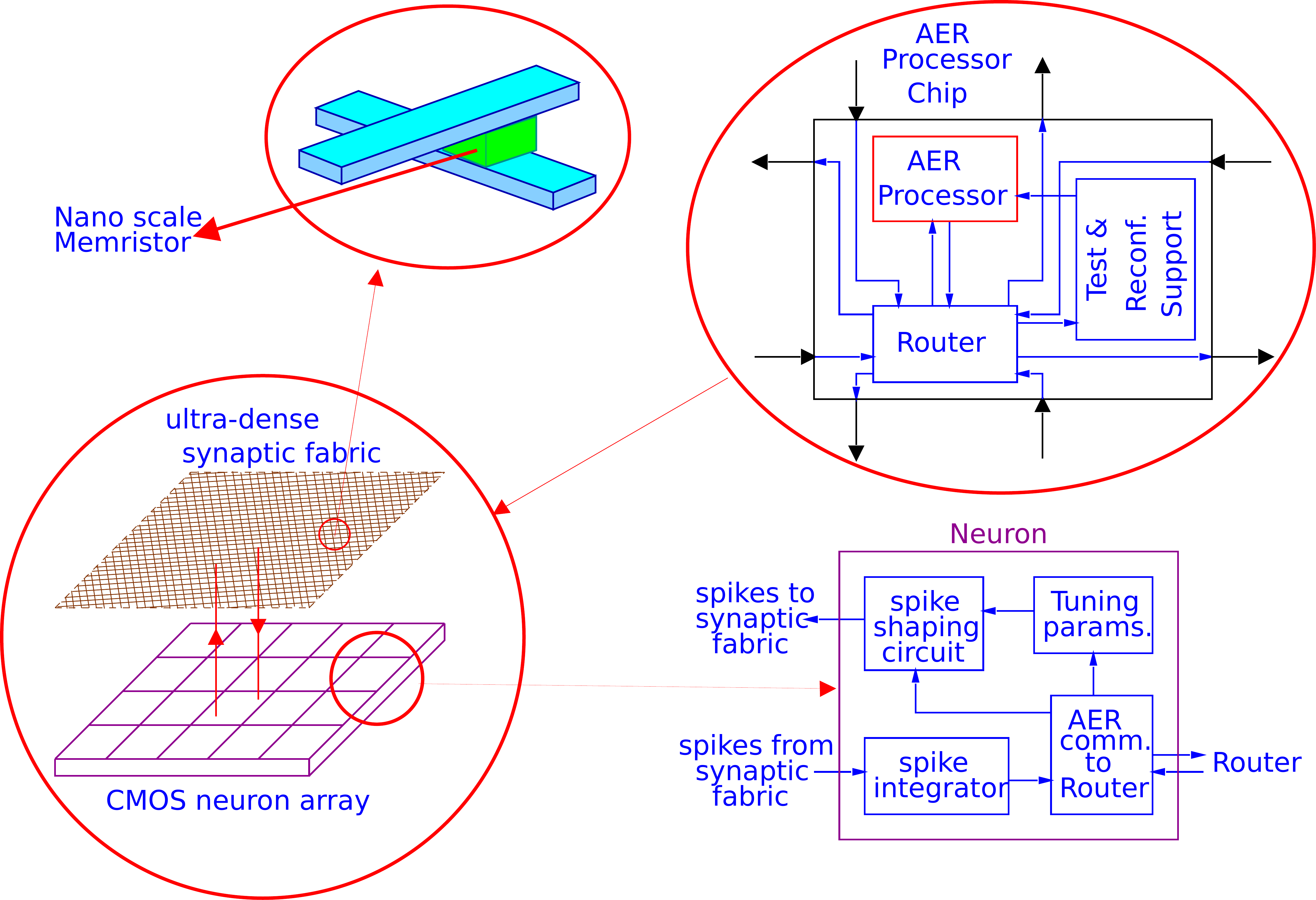}
    \caption{}
    \label{fig:system}
  \end{subfigure}
  \caption{Single memristor synapse concept.(a) One Memristor synapse
    with pre- and post-synaptic pulse-shaping neuron circuits. (b)
    Example of a \acs{STDP} weight update learning function
    $\xi(\Delta T)$, where $\Delta T$ represents the difference
    between the timing of the post-synaptic and pre-synaptic
    spikes. (c) Circuit architecture comprising three neuron layers
    connected by means of synaptic crossbars. (d) Hybrid
    memristor/CMOS neurons and \acs{AER} 2D chip architecture for
    spike/event routing and processing. Parts of this figure were
    adapted from~\cite{Serrano-Gotarredona_etal13}.}
    \label{fig:stdp_bernabe}
\end{figure}

In~\cite{zamarreno2011spike,Serrano-Gotarredona_etal13} the authors
proposed a scheme where neurons can drive memristive synapses to
implement a \ac{STDP}~\cite{Bi_Poo98} learning scheme by generating
single pairs of pre- and post-synaptic spikes in a fully asynchronous
manner, without any need for global or local synchronization, thus
solving some of the problems that existed with previously proposed
learning schemes~\cite{Snider2008,Kuzum:2012ge}.  The main idea is the
following: when no spike is generated, each neuron maintains a
constant reference voltage at both its input and output
terminals. During spike generation, each neuron forces a pre-shaped
voltage waveform at both its input and output terminals, as shown in
Fig.~\ref{fig:onesynapse}, to update the synaptic weight value stored
in the memristor state. Since memristors change their resistance when
the voltages at their terminals exceed some defined thresholds, it is
possible to obtain arbitrary \ac{STDP} weight update functions,
including biologically plausible ones, as the one shown in
Fig.~\ref{fig:stdpsim}~\cite{Bi_Poo98}. Moreover by properly shaping
the spike wave-forms of both pre- and post-synaptic spikes it is
possible to change the form of the \ac{STDP} learning function, or to
even make it evolve in time as learning
progresses~\cite{linares2009exploiting, zamarreno2011spike}. Fully
interconnected or partially interconnected synaptic crossbar arrays,
as illustrated in Fig.~\ref{fig:xbar}, could facilitate hierarchical
learning neural network architectures. Since there is no need for
global synchronization, this approach could be extended to multi-chip
architectures that transmit spikes across chip boundaries using fully
asynchronous timing. \revised{aer}{For example, a common asynchronous
  communication protocol that has been used in neuromorphic systems is
  based on the \ac{AER}~\cite{Deiss_etal98,Chicca_etal07b}}. In this
representation, each spiking neuron is assigned an address, and when
the neuron fires an address-event is put on a digital bus, at the time
that the spike is emitted. In this way time represent itself, and
information is encoded in real-time, in the inter-spike
intervals. 
By further exploiting hybrid \ac{CMOS}/memristor chip fabrication
techniques~\cite{Likharev_etal03}, this approach could be easily
scaled up to arbitrarily large networks (e.g., see
Fig.~\ref{fig:system}). Following this approach each neuron processor
would be placed in a 2D grid fully, or partially interconnected
through memristors. Each neuron would perform incoming spike
aggregation, provide the desired pre- and post-synaptic (programmable)
spike waveforms, and communicate incoming and outgoing spikes through
\ac{AER} communication circuitry. Using state-of-the-art CMOS
technology, it is quite realistic to provide in the order of a million
such neurons per chip with about $10^4$ synapses per neuron.
\revised{example}{For example, by using present day 40\,nm \ac{CMOS}
  technology it is quite realistic to fit a neuron within a $10\mu
  m\times 10\mu\, m$ area. This way, a chip of about $1cm^2$ could
  host of the order of one million neurons. At the same time, for the
  nano wire fabric deposited on top of \ac{CMOS} structures, present
  day technology can easily provide nano wires of 100nm
  pitch~\cite{Govoreanu:2011ij}. This would allow to integrate about
  $10^4$ synapses on top of the area occupied by each \ac{CMOS}
  neuron.} Similarly, at the \ac{PCB} level, it is possible to
envisage that a 100-chip PCB could host about $10^8$ neurons, and 40
of these PCBs would emulate 4 billion neurons. In these large-scale
systems the bottleneck is largely given by the spike or event
communication limits. To cope with these limits such chips would
inter-communicate through nearest neighbors, exploiting 2D-grid
network-on-chip (NoC) and network-on-board (NoB) principles. For
example, in~\cite{zamarreno2012AER} the authors proposed a very
efficient multi-chip inter-communication scheme that distributes event
traffic over a 2D mesh network locally within each board through
inter-chip high speed buses. Reconfigurability and flexibility would
be ensured by defining the system architecture and topology through
in-chip routing tables. Additionally, by arranging the neurons within
each chip in a local 2D mesh with in-chip inter-layer event
communication, it is possible to keep most of the event traffic inside
the chips. At the board level, the 2D mesh scheme would allow for a
total inter-chip traffic in the order of $E_v = 4 N_{ch}\times
E_{pp}$, where $N_{ch}=100$ is the number of chips per board, $E_{pp}$
is the maximum event bandwidth per inter-chip bus (which we may assume
to be around 100\,Meps - mega events per second), and 4 reflects the
fact that each chip is connected to its four nearest
neighbors~\cite{zamarreno2012AER}. With these numbers, the maximum
traffic per board would be in the order of $E_v\approx 4\times
10^{10}eps$, which is about 400\,eps per neuron just for inter-chip
event exchange. In practice, inter-board traffic could be much
sparser, if the system is partitioned efficiently. Such numbers are
quite realistic for present day CMOS technology, and the approach is
scalable. \revised{power}{Regarding power consumption of the
  communication overhead, we can use as reference some recent
  developments for event-based fully bit-serial inter-chip
  transmission schemes over differential micro
  strips~\cite{zamarreno2012LVDS,zamarreno2012AER}, where consumption
  is proportional to communication event rate. Each link would consume
  in the order of 40\,mA at 10\,Meps rate (this includes driver and
  receiver pad circuits~\cite{zamarreno2012LVDS} as well as
  serializers and deserializers~\cite{zamarreno2011SERDES}). If each
  neuron fires at an average rate of 1\,Hz, and if each chip has 1
  million neurons, the current consumption of the communication
  overhead would be about 4\,mA per chip. If voltage supply is in the
  1-2\,V range, this translates into 4-8\,mW per chip. For a 100 chip
  \ac{PCB} the inter-chip communication overhead power consumption
  would thus be about 400-800\,mW, for 1\,Hz average neuron firing
  rate.}

\section{Neuromorphic and hybrid memristor-CMOS synapse circuits}
\label{sec:synapse-circ}

\revised{real-synapses}{We've shown how memristive devices and
  nano-technologies can be exploited to dramatically increase
  integration density and implement large-scale abstract neural
  networks.  However to faithfully reproduce the function of real
  synapses, including their temporal dynamic properties, passive
  memristive devices would need to be interfaced to biophysically
  realistic \ac{CMOS} circuits that follow the neuromorphic approach,
  as described in~\cite{Indiveri_Horiuchi11,Mead90}. On one hand,
  building physical implementations of circuits and materials that
  directly emulate the biophysics of real synapses and reproduce their
  detailed real-time dynamics is important for basic research in
  neuroscience, on the other, this neuromorphic approach can pave the
  way for creating an alternative non-von~Neumann computing
  technology, based on massively parallel arrays of slow, unreliable,
  and highly variable, but also compact and extremely low-power
  solid-state components for building neuromorphic systems that can
  process sensory signals and interact with the user and the
  environment in real-time, and possibly carry out computation using
  the same principles used by the brain. Within this context, of
  massively parallel artificial neural processing elements, memory and
  computation are co-localized. Typically the amount of memory
  available per each ``computing node'' (synapse in our case) is
  limited and it is not possible to transfer and store partial results of a
  computation in large memory banks outside the processing
  array. Therefore,} in order to efficiently process real-world
biologically relevant sensory signals these types of neuromorphic
systems must use circuits that have biologically plausible time
constants (i.e.,\ of the order of tens of milliseconds). In this way,
in addition to being well matched to the signals they process, these
systems will also be inherently synchronized with the real-world
events they process and \revised{interact}{will be able to interact}
with the environment they operate in. But these types of time
constants require very large capacitance and resistance values. For
example, in order to obtain an equivalent $RC$ time constant of
$10$\,ms with a resistor even as large as $10$\,M$\Omega$, it would be
necessary to use a capacitor of $100$\,pF. In standard \ac{CMOS}
\ac{VLSI} technology a synapse circuit with this RC element would
require a prohibitively large area, and the advantages of large-scale
integration would vanish.  One elegant solution to this problem is to
use current-mode design techniques~\cite{Tomazou_Lidgey_etal90} and
log-domain \emph{subthreshold} circuits~\cite{Liu_etal02b,
  Mitra_etal10}. When \acp{MOSFET} are operated in the subthreshold
domain, the main mechanism of carrier transport is that of
diffusion~\cite{Liu_etal02b}, the same physical process that governs
the flow of ions through proteic channels across neuron membranes.  As
a consequence, \acp{MOSFET} have an exponential relationship between
gate-to-source voltage and drain current, and produce currents that
range from femto- to nano-Amp\`eres. In this domain it is possible to
implement active \ac{VLSI} analog filter circuits that have
biologically realistic time-constants and that employ relatively small
capacitors.

\subsection{A CMOS neuromorphic synapse}
\label{sec:cmos-neur-synapse}

\begin{figure}
  \centering
  \begin{subfigure}[b]{0.45\textwidth}
    \includegraphics[width=\textwidth]{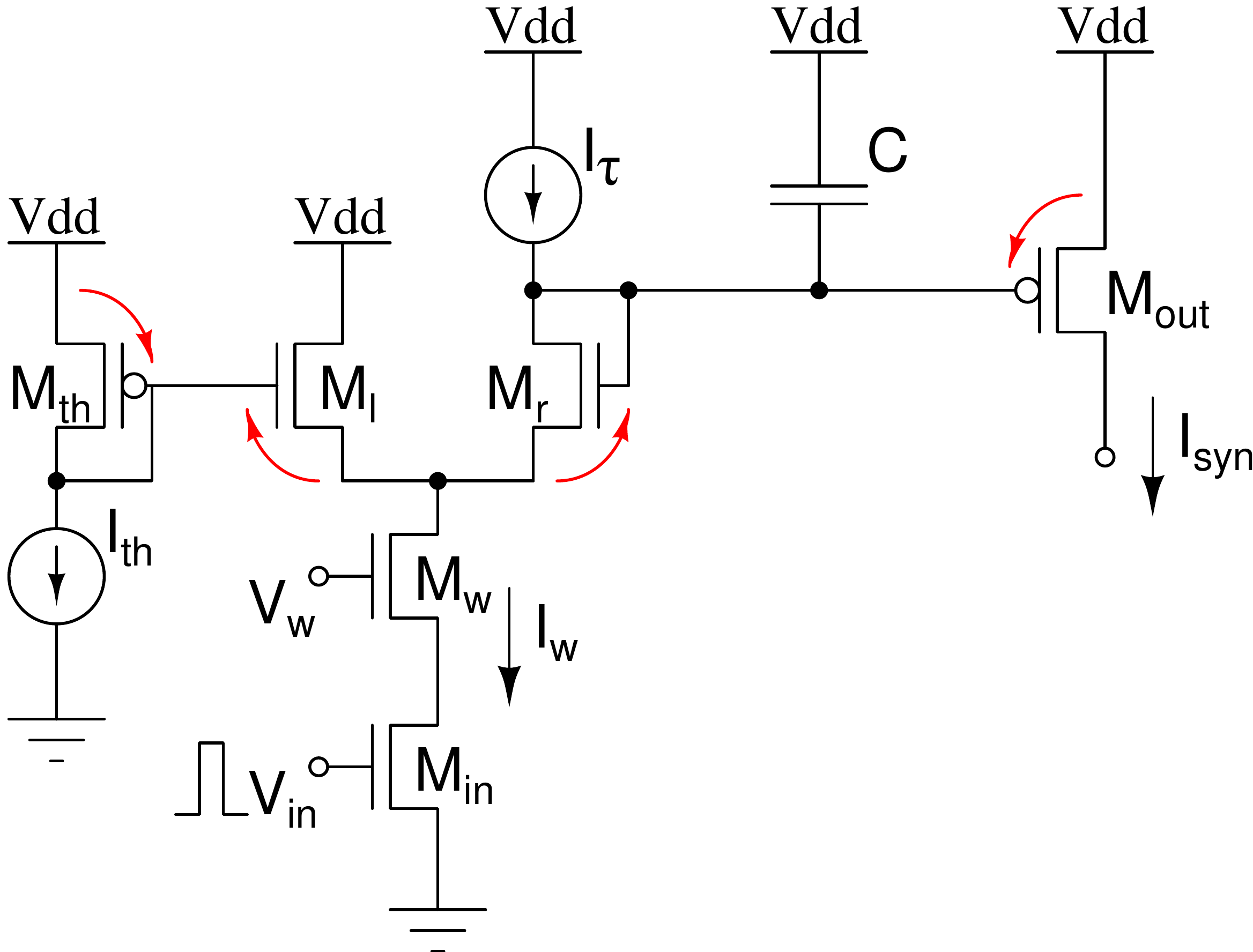}
    \caption{}
    \label{fig:dpi-circ}
  \end{subfigure}\hfill
  \begin{subfigure}[b]{0.5\textwidth}
    \includegraphics[width=\textwidth]{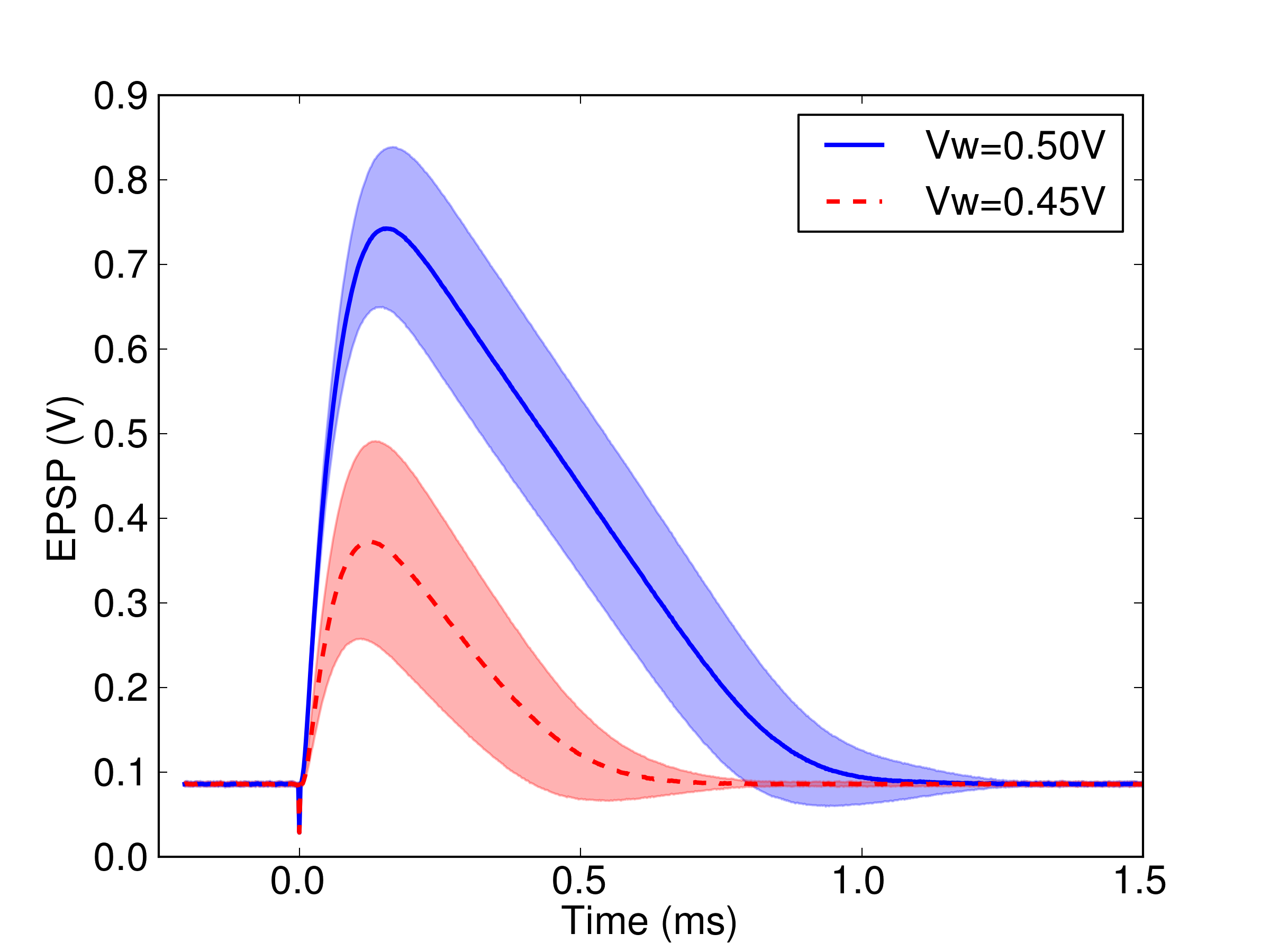}
    \caption{}
   \label{fig:dpi-epsp}
 \end{subfigure}
 \caption{Neuromorphic electronic synapses (a) Log-domain \acs{DPI}
   circuit diagram of an excitatory silicon synapse. Red arrows show
   the translinear loop considered to derive the circuit
   response. Input voltage spikes $V_{in}$ are integrated by the
   circuit to produce post-synaptic currents $I_{syn}$ with
   biologically faithful dynamics. (b) Experimental data showing the
   \acs{EPSP} response of the circuit for two different settings of
   synaptic weight bias voltage $V_w$. The data was measured from the
   \acs{DPI} synapses of 124 neurons, integrated on the same chip, with
   shared common bias settings. The dashed and solid lines represent
   the average response, while the shaded areas (standard deviation)
   indicate the extend of the device mismatch effect.}
    \label{fig:dpi}
\end{figure} 

An example of a compact circuit that can produce both linear dynamics
with biologically plausible time constants as well as non-linear
short-term plasticity effects analogous to those observed in real
neurons and synapses is \revised{acronym1}{the \ac{DPI}
circuit~\cite{Bartolozzi_Indiveri07b}} shown in
Fig.~\ref{fig:dpi-circ}.  It can be shown~\cite{Bartolozzi_etal06}
that by exploiting the \emph{translinear-principle}~\cite{Gilbert96}
across the loop of gate-to-source voltages highlighted in the figure,
the circuit produces an output current $I_{syn}$ with impulse response
of the form:
\begin{equation}
  \label{eq:linear}
  \tau \frac{d}{dt}I_{syn} + I_{syn} = \frac{I_{w}I_{th}}{I_{\tau}},
\end{equation}
where $\tau \triangleq {C U_{T}}/{\kappa I_{\tau}}$ is the circuit
time constant, $\kappa$ the subthreshold slope
factor~\cite{Liu_etal02b}, and $U_{T}=KT/q$ represents the thermal
voltage. The currents $I_{w}$ and $I_{th}$ represent local synaptic
weight and a global synaptic scaling gain terms, useful for
implementing spike-based and homeostatic plasticity
mechanisms~\cite{Abbott_Nelson00,Turrigiano_Nelson04}.
Therefore, by setting for example, $I_{\tau}=5\,p$A, and assuming that
$U_{T}=25$\,mV at room temperature, the capacitance required to
implement a time constant of $10$\,ms would be approximately
$C=1$\,pF.  This can be implemented in a compact layout and allows the
integration of large numbers of silicon synapses with realistic
dynamics on a small \ac{VLSI} chip.  The same circuit of
Fig.~\ref{fig:dpi-circ} can be used to implement elaborate models of
spiking neurons, such as the ``Adaptive Exponential'' (AdExp) \ac{IF}
model~\cite{Brette_Gerstner05,Indiveri_etal11}.  Small (minimum-size,
of about $10\,\mu$m$^2$) prototype \ac{VLSI} chips comprising of the
order of thousands of neurons and synapses based on the \ac{DPI}
circuit have been already fabricated using a conservative 350\,nm
\ac{CMOS} technology~\cite{Indiveri_Chicca11}. The data of
Fig.~\ref{fig:dpi-epsp} shows the average response of a \ac{DPI}
synapse circuits measured from one of such
chips~\cite{Indiveri_Chicca11}. The data represents the average
\ac{EPSP} produced by 124 neurons in response to a single spike sent
to the \ac{DPI} synapses of each neuron.  The shaded areas,
representing the standard deviation, highlight the extent of
variability present in these types of networks, due to device
mismatch. The main role of the \ac{DPI} circuit of
Fig.~\ref{fig:dpi-circ} is to implement synaptic dynamics. Short-term
plasticity, \ac{STDP} learning, and homeostatic adaptation mechanisms
can be, and have been, implemented by interfacing additional \ac{CMOS}
circuits to control the \ac{DPI} $V_w$ bias voltage, or to the
$I_{th}$ bias
current~\cite{Bartolozzi_Indiveri07b,Mitra_etal09,Bartolozzi_Indiveri09}. Long-term
storage of the $V_w$ weights however requires additional
power-consuming and area-expensive circuit solutions, such as floating
gate circuits, or local \ac{ADC} and \ac{SRAM} cells.

\subsection{A new hybrid memristor-CMOS neuromorphic synapse}
\label{sec:hybr-memr-neur}

\begin{figure}
  \centering
  \begin{subfigure}[b]{0.375\textwidth}
    \includegraphics[width=\textwidth]{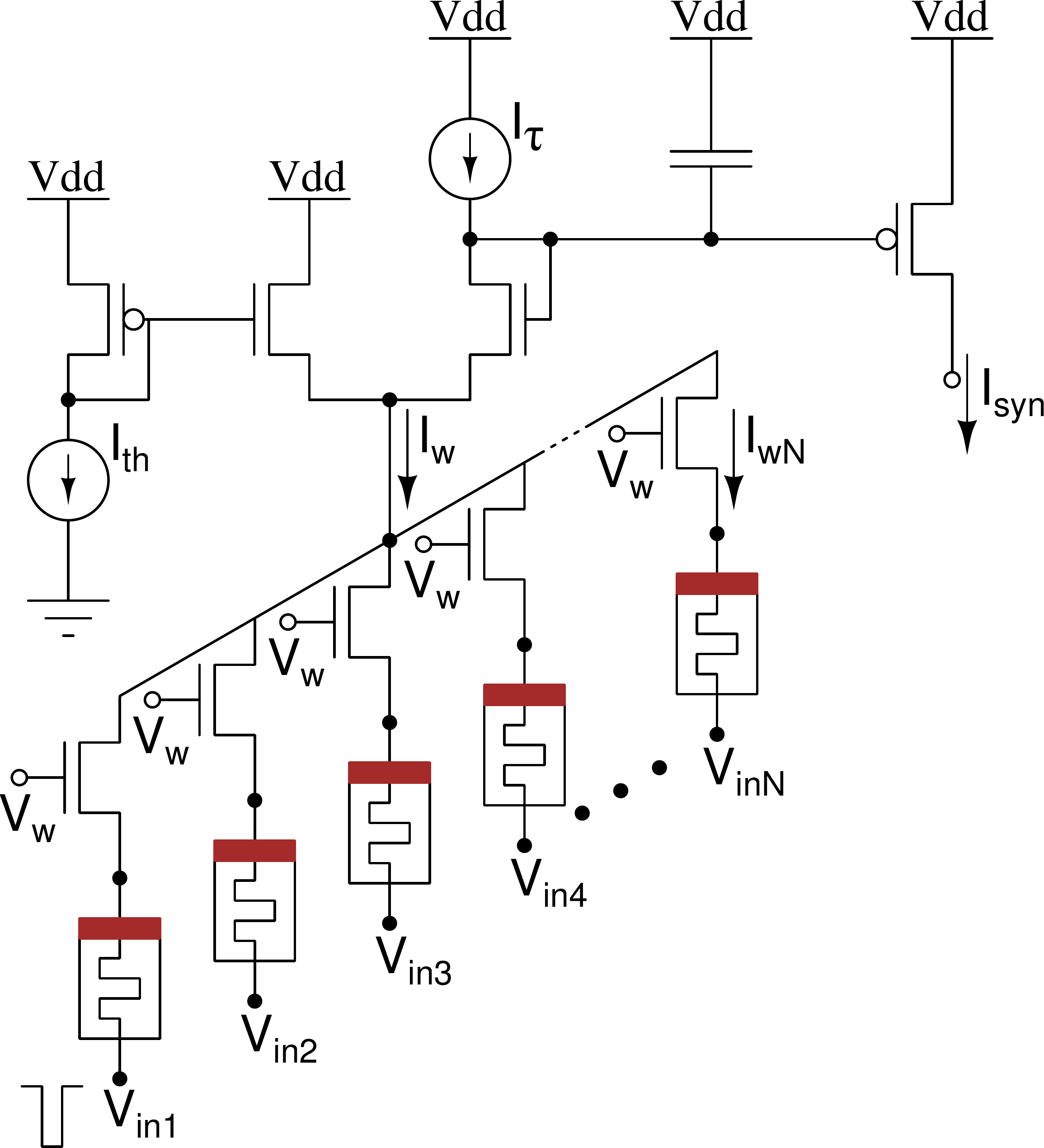}
    \caption{}
    \label{fig:nano-synapse}
  \end{subfigure}\hfill
  \begin{subfigure}[b]{0.525\textwidth}
    \includegraphics[width=\textwidth]{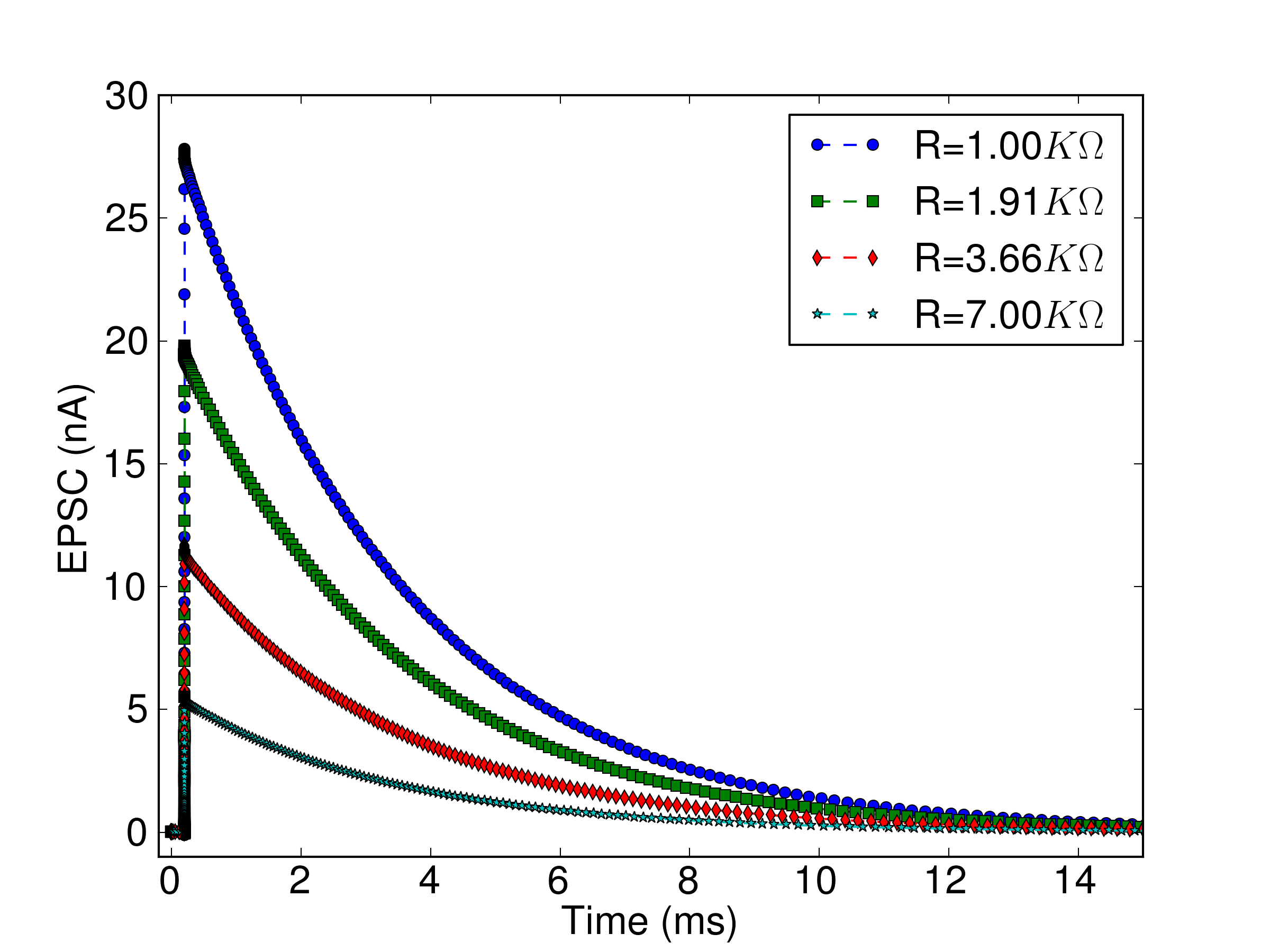}
    \caption{}
    \label{fig:nano-synapse-sim}
  \end{subfigure}
  \caption{Neuromorphic memristive synapse. (a) Schematic circuit
   implementing an array of memristive synapses, with independent
   inputs and synaptic weights, but with shared temporal dynamics. (b)
   SPICE simulations of the circuit in Fig.~\ref{fig:nano-synapse} showing
   the output $I_{syn}$ \acs{EPSC} in response to a pre-synaptic input
   spike, for 4 different memristor conductance values.}
  \label{fig:dpimemr}
\end{figure}

\revised{propose}{Nano-electronic technologies offer a promising
  alternative solution for compact and low-power long-term storage of
  synaptic weights. The hybrid memristor-CMOS neuromorphic synapse
  circuit we propose here, shown in Fig.~\ref{fig:nano-synapse},
  exploits these features to obtain at the same time dense integration
  of low-power long-term synaptic weight storage elements, and to
  emulate detailed synaptic biophysics for implementing relevant
  computational properties of neural systems.}

\revised{stochastic}{The circuit depicted in
  Fig.~\ref{fig:nano-synapse} represents a possible implementation of
  a dense array of N synapses with independent weights but with the
  same, shared, temporal dynamics. Depending on their size, each
  memristor in Fig.~\ref{fig:nano-synapse} could represent a full
  synaptic contact, or an individual ion channel in the synaptic cleft
  (see also Fig.~\ref{fig:synapse}).  If the currently accepted model
  of filament formation in memristive devices is true, then
  down-scaled memristors should approach single filament bistable
  operation.  While this is a severe limitation for classical neural
  network applications in which memristors are required to store
  analog synaptic weight values with some precision, it would actually
  provide a very compact physical medium for emulating the stochastic
  nature of the opening and closing of ion channels in biological
  synapses.}

The shared temporal dynamics are implemented by the \ac{DPI} circuit
in the top part of Fig.~\ref{fig:nano-synapse}. Indeed, if this
circuit is operated in its linear regime, it is possible to
time-multiplex the contributions from all spiking inputs, thus
requiring one single integrating element and saving precious silicon
real-estate.  The $V_w$ bias voltage of this circuit is a global
parameter that sets the maximum possible current that can be produced
by each memristor upon the arrival of an input spike, while the
memristor conductance modulates the current being produced by the
synapse very much like conductance changes in real synapses affect
\revised{acronym2}{the \acp{EPSC}} they produce. Larger memristor
conductances, which represent a larger number of open proteic channels
in real synapses, correspond to larger synaptic weights. 

Figure~\ref{fig:nano-synapse-sim} shows the results of SPICE
simulations of the circuit in Fig.~\ref{fig:nano-synapse}, for a
180\,nm CMOS process. The $I_{thr}$ and $I_{\tau}$ current sources
were implemented with p-type \acp{MOSFET}, biased to produce 2\,pA and
10\,pA respectively, and the $V_w$ voltage bias was set to
700\,mV. \revised{newdata}{The data was obtained by simulating the
  response of one input memristive branch to a single input spike,
  while sweeping the memristor impedance from 1\,K$\Omega$ to
  7\,K$\Omega$. In these simulations we set the memristor in its
  \ac{LRS}, and assumed we could modulate the value of the resistance
  to obtain four distinct analog states analogous to the ones measured
  experimentally in Fig.~\ref{fig:mempulsing}. Of course the circuit
  supports also the operation of the memristor as a binary device,
  working in either the \ac{HRS} state or the \ac{LRS} one. This
  bi-stable mode of using the memristor would encode only an ``on'' or
  ``off'' synaptic state, but it would be more reliable and it} is
compatible with biologically plausible learning mechanisms, such as
those proposed in~\cite{Brader_etal07}, and implemented
in~\cite{Mitra_etal09}.  The circuit of Fig.~\ref{fig:nano-synapse}
shows only the circuit elements required for a ``read'' operation,
i.e.,\ an operation that stimulates the synapse to generate an
\ac{EPSC} with an amplitude set by the conductance of the
memristor. Additional circuit elements would be required to change the
value of the memristor's conductance, e.g., via learning
protocols. However the complex circuitry controlling the learning
mechanisms would be implemented at the \ac{IO} periphery of the
synaptic array, for example with pulse-shaping circuits and
architectures analogous to the ones described in
Section~\ref{sec:cmos-nano}, or with circuits that check the state of
the neuron and of it's recent spiking history, such as those proposed
in~\cite{Mitra_etal10}, and only a few additional compact elements
would be required in each synapse to implement the weight update
mechanisms.

\section{Brain-inspired probabilistic  computation}
\label{sec:brain-insp-comp}

While memristors offer a compact and attractive solution for long-term
storage of synaptic state, as done for example in
Fig.~\ref{fig:dpimemr}, they are affected by a high degree of
variability (e.g., much higher than the one measured for \ac{CMOS}
synapses in Fig.~\ref{fig:dpi-epsp}). In addition, as memristors are
scaled down, unreliable and stochastic behavior becomes unavoidable.
The variability, stochasticity, and in general reliability issues that
are starting to represent serious limiting factors for advanced
computing technologies, do not seem to affect biological computing
systems. Indeed, the brain is a highly stochastic system that operates
using noisy and unreliable nanoscale elements. Rather than attempting
to minimize the effect of variability in nanotechnologies, one
alternative strategy, compatible with the neuromorphic approach, is to
embrace variability and stochasticity and exploit these ``features''
to carry out robust brain-inspired probabilistic computation.

The fact that the brain can efficiently cope with a high degree of
variability is evident at many levels: at the macroscopic level
trial-to-trial variability is present for example in the arm
trajectories of reaching movement tasks. It is interesting to note
that the variability of the end position of the reaching movement is
reduced, if the task requires to hit or touch a target with high
accuracy~\cite{TodorovJordan:02}. Variability is evident at the level
of cortical neurons: there is significant trial to trial variability
in their responses to identical stimuli; it is evident also at the
level of chemical synapses, where there is a high degree of
stochasticity in the transmission of neurotransmitter
molecules~\cite{FaisalETAL:08}, from the pre-synaptic terminal to the
post-synaptic one.  The release probability of cortical synapses
ranges from values of less than 1\% to
100\%~\cite{volgushev2004probability}.  This indicates that stochastic
synaptic release may not merely be an unpleasant constraint of the
molecular machinery but may rather be an important computational
feature of cortical synapses.


What could be the computational benefit of using hardware affected by
variability and stochasticity in biological and artificial computing
systems?  Recent advances in cognitive science demonstrated that human
behavior can be described much better in the framework of
\emph{probabilistic inference} rather than in the framework of
traditional ``hard'' logic inference~\cite{GriffithsTenenbaum:06}, and
encouraged the view that neuronal networks might directly implement a
process of probabilistic inference~\cite{Tenenbaum2011_how}.
\revised{inference}{In parallel, to this paradigm shift, research in
  machine learning has revealed that probabilistic inference is often
  much more appropriate for solving real-world problems, then hard
  logic~\cite{Pearl:88}. The reason for this is that reasoning can
  seldom be based on full and exact knowledge in real-world
  situations. For example, the sensory data that a robot receives is
  often noisy and incomplete such that the current state of the
  environment can only partially be described. Probabilistic reasoning
  is a powerful tool to deal with such uncertain situations. Of
  course, exact probabilistic inference is still computationally
  intractable in general, but a number of approximation schemes have
  been developed that work well in practice.}

In probabilistic inference, the idea is to infer a set of unobserved
variables (e.g., motor outputs, classification results, etc.)  given a
set of observed variables (evidence, e.g., sensory inputs), using
known or learned probabilistic relationships among them.
Specifically, if the distribution $P(\bar{x})$ describes the
probabilistic relationships between the random variables $x_1, \dots,
x_n$, and if $x_1, \dots ,x_k$ of this distribution are observed, then
one can infer a set of variables of interests $x_{k+1}, \dots,
x_{k+l}$ by determining the posterior probability $P(x_{k+1}, \dots,
x_{k+l}|x_1, \dots x_k)$.
One of the most popular techniques used to perform inference is
\emph{belief propagation}~\cite{Pearl:88}. While this message passing
algorithm can be implemented by networks of spiking
neurons~\cite{Steimer2009_belief}, a more promising alternative
approach, also well suited to model brain-inspired computation, is
to use \emph{sampling techniques}~\cite{BuesingETAL:11}.
Probably the most important family of sampling techniques in this
context is \ac{MCMC} sampling. Since \ac{MCMC} sampling techniques
operate in a stochastic manner, stochastic computational elements are
a crucial and essential feature.  Recent studies have shown that
probabilistic inference through \ac{MCMC} sampling can be implemented
by networks of stochastically spiking neurons~\cite{BuesingETAL:11,
  PecevskiETAL:11a}. Therefore, \ac{MCMC} sampling is a computational
paradigm optimally suited for emulating probabilistic inference in the
brain using neuromorphic circuits and nanoelectronic synapses.

Within this context, it is important to see if and how the
distribution $P(\bar{x})$ can be \emph{learned} from observations,
i.e.,\ how the artificial neural system can build its own model of the
world based on its sensory input and then perform probabilistic
inference on this model. For a relatively simple
model~\cite{NesslerETAL:10}, it has been shown that this can be
accomplished by a local spike-driven learning rule that resembles the
\ac{STDP} mechanisms measured in cortical
networks~\cite{Bi_Poo98}. Analogous learning mechanisms have been
demonstrated both experimentally in neuromorphic \ac{CMOS}
devices~\cite{Mitra_etal09}, and theoretically, with circuit models of
memristive synapses~\cite{zamarreno2011spike}.

With regard to learning, the variability and stochasticity
``features'' described above can provide an additional benefit: for
many learning tasks, humans and animals have to explore many different
actions in order to be able to learn appropriate responses in a given
situation. In these so-called reinforcement learning setups, noise and
variability naturally provide the required exploration mechanisms. A
number of recent studies have shown how stochastic neuronal behavior
could be utilized by cortical circuits in order to learn complex
tasks~\cite{LegensteinETAL:08a,LegensteinETAL:09b,HoerzerETAL:12}.
For example, Reservoir Computing (RC, also known under the terms
Liquid State Machines and Echo State Networks) is a powerful general
principle for computation and learning with complex dynamical systems
such as recurrent networks of analog and spiking neurons
\cite{Jaeger2004,Maass2002a} or optoelectronic devices
\cite{Paquot:2012}. The main idea behind RC is to use a heterogeneous
dynamical system (called the reservoir) as a nonlinear fading memory
where information about previous inputs can be extracted from the
current state of the system.  This reservoir can be quite arbitrary in
terms of implementation and parameter setting as long as it operates
in a suitable dynamic regime \cite{Legenstein2007}.  Readout elements
are trained to extract task-relevant information from the
reservoir. In this way, arbitrary fading memory filters or even
arbitrary dynamical systems (in the case when the readout elements
provide feedback to the dynamical system) can be learned. One
long-standing disadvantage of traditional RC was that readouts had to
be trained in a supervised manner. In other words, a teacher signal
was necessary that signals at each time point the desired output of
readouts. \revised{teacher}{In many real-world applications, such a
  teacher signal is not available. For example, if the task for a
  robot controller is to produce some motor trajectory in order to
  produce a desired hand movement, the exact motor commands that
  perform this movement are in general not known. What can be
  evaluated however is the quality of the movement.} Recently, it has
been demonstrated that noisy readouts can be trained with a much less
informative reward signal, which just indicates whether some measure
of performance of the system has recently
increased~\cite{HoerzerETAL:12}. \revised{slow-learning}{Of course,
  such reward-based learning can in general be much slower than the
  pure supervised approach (see, e.g.,\cite{Urbanczik_Senn09}). The
  actual slowdown however depends on the task at hand, and it is
  interesting that for a set of relevant tasks, reward-based learning
  works surprisingly fast~\cite{HoerzerETAL:12}.}

Since the functionality of reservoirs depends on its general dynamical
behavior and not on precise implementation of its components, RC is an
attractive computational paradigm for circuits comprised of nanoscale
elements affected by variability, such as the one proposed in
Section~\ref{sec:hybr-memr-neur}.  In fact, if the reservoir is
composed by a large number of simple interacting dynamic elements --
the typical scenario -- then heterogeneity of these elements is an
essential requirement for ideal performance.  Parameter heterogeneity
is also beneficial in so-called ensemble learning techniques
\cite{Rokach:2010}. It is well-known that the combination of models
with heterogeneous predictions for the same data-set tends to improve
overall prediction performance \cite{Kuncheva:2003}. Hence,
heterogeneity of computational elements can be a real benefit for
learning. Examples for ensemble methods are random forests
\cite{Breiman:2001}, bagging \cite{Breiman:96}, and boosting
\cite{Schapire:1990}.

\section{Discussion and conclusions}
\label{sec:discussion}

Memristors, and in particular nanoscale solid state implementations,
represent a promising technology, baring benefits for emerging memory
storage as well as revisiting conventional analog
circuits~\cite{berdan2012memristive}.  Given their low-power and
small-scale characteristics, researchers are considering their
application also in large-scale neural networks for neuro-computing
applications.  However, the fabrication of large-scale nano-scale
cross-bar arrays involves several issues that are still open:
the realization of nano sized electrodes requires
nanopatterning~\cite{xia_self-aligned_2010} techniques, such as
\ac{EBL} or \ac{NIL}~\cite{xia_nanoscale_2011}. This directly
correlates to reduced electrode cross section which results in
increasing resistance. As electrode resistance scales with length,
this can rapidly become a critical issue for fully interconnected
nanoscale cross-bar structures. Furthermore, down-scaling the electrode
size to reduce the device active area requires simultaneous
down-scaling of the thickness of the metalizations due to fabrication
concerns. This in turn further increases the resistance of the
electrodes, much like the interconnects in modern \ac{CMOS}
circuitry. These factors introduce a large offset in the write
voltages required to change the state of \acp{RRAM} cells that depends
on the position of the cell in the array.  This problem is especially
critical in neuro-computing architectures where these cells represent
synapses, as the offsets directly affect the weight-update and
learning mechanisms.

Integrating memristors as synapse elements in large-scale
neuro-computing architectures also introduces the significance of
process variability in memristor dimensions~\cite{hu_geometry_2011},
which in turn introduces a significant amount of variability in the
characteristics of the synapse properties.
In addition to their large variability, another important issue
relating to these types of synapses, that is still ignored in the vast
majority of neuro-computing studies, is the effect of limited
resolution in memristive states. In particular, it is not known what
the trade-off between desired synaptic weight resolution and memristor
size is. And it is not known to what extent the multi-step synaptic
weight model holds true for aggressively down-scaled memristor
sizes.

These scaling, integration, and variability issues are serious
limiting factors for the use of memristors in conventional
neuro-computing architectures. Nonetheless, biological neural systems
are an existence proof that it is possible to implement robust
computation using nanoscale unreliable components \revised{nonvN}{and
  non-von~Neumann computing architectures}.  In order to best exploit
these emerging nanoscale technologies for building compact, low-power,
and robust artificial neural processing systems it is important to
understand the (probabilistic) neural and cortical principles of
computation and to develop at the same time, following a co-design
approach, the neuromorphic hardware computing substrates that support
them.
In this paper we elaborated on this neuromorphic approach, presenting
an example of a neuromorphic circuit and of a hybrid
nanoelectronic-\ac{CMOS} architecture that directly emulate the
properties of real synapses to reproduce biophysically realistic
response properties, thus providing the necessary technology for
implementing massively parallel models of brain-inspired computation
that are, by design, probabilistic, robust to variability, and fault
tolerant.




\section*{Acknowledgment}

This work was supported by the European CHIST-ERA program, via the
``Plasticity in NEUral Memristive Architectures'' (PNEUMA) project.

\section*{References}
\bibliographystyle{unsrt}
\bibliography{nano12-memristors}

\end{document}